\documentclass{aa}

\usepackage[dvipsnames]{xcolor} 
\usepackage{graphicx, subfigure, amsmath, pifont}
\usepackage{amssymb}
\usepackage{multicol}
\usepackage[breaklinks=true]{hyperref} 
\hypersetup{colorlinks=true,citecolor=Blue}
\usepackage{natbib}
\bibpunct{(}{)}{;}{a}{}{,} 
\usepackage[english]{babel}
\usepackage{blindtext}
\usepackage[normalem]{ulem}
\usepackage{comment}
\usepackage{lmodern}
\usepackage{ulem} 
\usepackage{soul}

\let\oldpageref\pageref
\renewcommand{\pageref}{\oldpageref*}

\graphicspath{{./}{figures/}}
\graphicspath{{./}{figures/FARGO/}}

\begin{document} 

 \title{Recovering the gas properties of protoplanetary disks through parametric visibility modeling: MHO\,6}

   \author{
    N.~T.~Kurtovic\inst{\ref{mpia}},
    P.~Pinilla\inst{\ref{uclondon}},
   }
   \institute{
   Max-Planck-Institut f\"{u}r Astronomie, K\"{o}nigstuhl 17, 69117, Heidelberg, Germany, \email{nicokurtovic@gmail.com} \label{mpia}
   \and Mullard Space Science Laboratory, University College London, Holmbury St Mary, Dorking, Surrey RH5 6NT, UK. \label{uclondon}
   }
   \date{}

 \authorrunning{N.T.~Kurtovic}
 \titlerunning{Very Low Mass Star: MHO\,6}

  \abstract
   {The composition and distribution of the gas in a protoplanetary disk plays a key role in shaping the outcome of the planet formation process. Observationally, the recovery of information such as the emission height and brightness temperature from interferometric data is often limited by the imaging processes.} 
   {To overcome the limitations of image-reconstruction when analyzing gas emission from interferometric observations, we have introduced a parametric model to fit the main observable properties of the gaseous disk component in the visibility plane. This approach is also known as parametric visibility modeling. } 
   {We applied our parametric visibility modeling to the gas brightness distribution of the molecular line emission from $^{12}$CO J=3-2 and $^{13}$CO J=3-2 in the disk around MHO\,6, a very-low-mass star in the Taurus star-forming Region. To improve the flux fidelity of our parametric models, we combined models with different pixel resolution before the computation of their visibilities, referred to as ``nesting images.'' } 
   {When we apply our parametric visibility modeling to MHO\,6, with independent fits to the emission from its CO isopotologues, the models return the same consistent results for the stellar mass, disk geometry, and central velocity. The surface height and brightness temperature distribution are also recovered. When compared to other disks, MHO\,6 surface height is among the most elevated surfaces, consistent with the predictions for disks around very-low-mass stars. } 
   {This work demonstrates the feasibility of running rapidly iterable parametric visibility models in moderate resolution and sensitivity interferometric observations. More importantly, this methodology opens the analysis of disk's gas morphology to observations where image-based techniques are unable to robustly operate, as in the case of the compact disk around MHO\,6. }
  

  \maketitle

\section{Introduction}\label{sec:intro}

The gas emission of protoplanetary disks is a key observational tracer of the temperature distribution \citep[e.g.,][]{schwarz2016, zhang2021}, kinematics \citep[e.g.,][]{teague2019, wolfer2023}, morphology of the emitting region \citep[e.g.,][]{pinte2018, Law2021, paneque2023}, and chemical properties \citep[e.g.,][]{Oberg2021}. All of these properties play a key role in shaping the architecture and composition of forming planets. 
The majority of these studies have only become possible through the superb sensitivity and angular resolution of interferometers such as the Atacama Large (sub-)Millimeter Array (ALMA). 
Interferometers sample the Fourier transform of the observed source, also known as the ``visibilities''  \citep[for a more detailed explanation, we refer to][and references therein]{czekala2021}, and images of the brightness distribution need to be reconstructed from the visibilities with algorithms such as CLEAN \citep{hogbom1974, clark1980, briggs1995} or maximum entropy and likelihood methods \citep[e.g.,][]{ponsonby1973, ables1974, cornwell1985, narayan1986, casassus2006, carcamo2018}. Even though the visibility coverage of ALMA is dense enough to recover high fidelity reconstructed images of its observations, any analysis performed on those images will be dependent on the specific imaging method chosen, as different images can deviate with regard to their features \citep[see ][]{zawadzki2023}. 

In the case of low signal-to-noise ratio (S/N) observations, information recovery from interferometric data can be maximized when assuming that the brightness distribution follows a coherent morphology. This is a common assumption for protoplanetary disks, which have a dominant axisymmetric structure \citep[e.g.,][]{andrews2018, long2018, cieza2021}. In the dust continuum emission, applying priors on the morphology of disks allows for parametric \citep[e.g.,][]{Tazzari2017} and non-parametric \citep[e.g.,][]{jennings2020} models to successfully recover very small-scale structures \citep[e.g.,][]{Kurtovic2022, jennings2022}, to search for faint emitting sources \citep[e.g.,][]{andrews2021}, and to characterize disk sizes and brightness distribution \citep[e.g.,][]{hendler2020}. The visibility modeling of the gas emission, however, has the additional challenge of generating velocity-dependent models to follow the disk's rotation, which considerably increases the difficulty of implementing prior emission morphologies compared to the dust visibility modeling. 
Nevertheless, visibility-based approaches to analyzing the gas emission of protoplanetary disks have been applied when recovering the turbulence level of disks \citep[e.g.,][]{flaherty2015, flaherty2020} and the dynamical mass of stellar hosts \citep{czekala2015, pegues2021, long2021}, as well as in match-filtering faint emission \citep{loomis2018}.

In this work, we explore the advantages and limitations of describing the properties of the gas surface-emitting layer with a parametric visibility-based modeling. Our aim is to extend the tools that are widely used to analyze dust continuum emission to the gas emission observations. 
We analyze the source MHO\,6, an M5.0 star a mass of $M_\star\sim0.2\,M_\odot$ \citep{kurtovic2021, pegues2021}, located in the Taurus star-forming region at 154\,pc \citep[Gaia DR3,][]{gaia2021edr3}. 
MHO\,6 was chosen as it is spatially resolved in dust continuum and $^{12}$CO J=3-2 emission when observed with 15\,au resolution. Also, it is one of the few disks around a very low mass star (VLMS) with a large gap at millimeter wavelengths \citep{pinilla2022}, which could have been produced by planet-disk interaction  \citep{kurtovic2021}. 
In the 0.9\,mm continuum, MHO\,6 shows a single ring peaking at 11\,au; while in the gas emission, the disk extends for approximately 200\,au in radius. Unlike several other VLMS disks in the Taurus Region, the CO isotopologue emission in the MHO 6 disk is not completely absorbed by the cloud in the low-velocity channels \citep[e.g., ][]{kurtovic2021, hashimoto2021}. This allows us to detect the disk in blueshifted and redshifted molecular line emission, with some absorption detected close to the systemic velocity of MHO\,6. 
Due to its moderate brightness and relatively small angular extent, MHO\,6 is not an ideal source to be studied in the image plane, as it would need to be targeted with very extended baselines for long exposure times. For these reasons, we chose to describe the disk morphology directly in the visibilities. 

In Sect.~\ref{sec:observations}, we summarize the observations analyzed in this work. In Sect.~\ref{sec:methodology} we describe the visibility fitting process, and we introduce our solution for generating intensity models of the inner disk emission. We present the results of our modeling in Sect.~\ref{sec:results}, and we further discuss the benefits and limitations of our visibility approach in Sect.~\ref{sec:discussion}. Finally, we present our conclusions in Sect.~\ref{sec:conclusions}.

\section{Observations}\label{sec:observations}

This work includes 0.87\,mm observations of the disk around MHO\,6. The dataset was obtained by ALMA in Band 7, as part of the project 2018.1.00310.S (PI: P.~Pinilla). The self-calibration and data preparation process was already published in \citet{kurtovic2021}. As a summary, our observations include two spectral windows covering dust continuum emission: one spectral window covering the molecular line $^{12}$CO J=3-2 transition (from now on referred to as $^{12}$CO) and one spectral window covering the molecular line $^{13}$CO J=3-2 transition (from now on referred to as $^{13}$CO). With \texttt{CASA 5.6.2} \citep{mcmullin07}, we averaged the continuum-subtracted measurement sets into $30$\,s bins to reduce data volume. We applied the task \texttt{cvel2} to the self-calibrated and time-binned $^{12}$CO, redefining the channels to start from -3.5\,km\,s$^{-1}$, with 0.5\,km\,s$^{-1}$ of channel width to increase the S/N of the emission, using 32 channels in total. 
This range covers the channels where emission was detected, along with a few channels with undetected emission on each side of the line. Compared to the native frequency resolution of the observation, we increased the channel width from 244.1\,kHz to 576.7\,kHz. For the $^{13}$CO, we started from -3.5\,km\,s$^{-1}$ with 0.9\,km\,s$^{-1}$ of channel width, using 19 channels in total to cover a similar velocity range as the $^{12}$CO. 
The visibility table of each channel was extracted and converted to wavelength units using each channel central frequency. 

For visual reference, we generated image cubes of each line with the CASA task \texttt{tclean}, with a robust parameter of 0.5 to obtain a good balance between the sensitivity and the angular resolution. No visibility tapering was applied. The beam size for the $^{12}$CO cube is $110\times80$\,mas with a position angle of $31.5\,$deg. The $^{13}$CO cube has a resolution of $124\times89$\,mas with a position angle of  $34.6\,$deg. Thus, the spatial resolution of each image cube is about $16\times12$\,au and $18\times13$\,au for $^{12}$CO and $^{13}$CO, respectively. Additional image cubes from synthetic models or residuals were reconstructed with the same \texttt{tclean} setup, and they share the observations angular resolution.


\section{Methodology: Parametric visibility modeling}\label{sec:methodology}

We analyze the $^{12}$CO and $^{13}$CO emission of MHO\,6 directly in the visibility plane. To do so, we followed a similar approach as that applied to the continuum visibility modeling presented in \citet{Kurtovic2022}. We generated a model image following a parametric description for the intensity distribution, and we used the \texttt{galario} package \citep{Tazzari2017} to calculate the visibilities of the model. When fitting gas emission, an image has to be generated for each fitted channel, as the brightness distribution is velocity-dependent. Therefore, we generated 32 images to fit the $^{12}$CO and 19 for the $^{13}$CO.

\begin{figure}[t]
\centering
    \includegraphics[width=9cm]{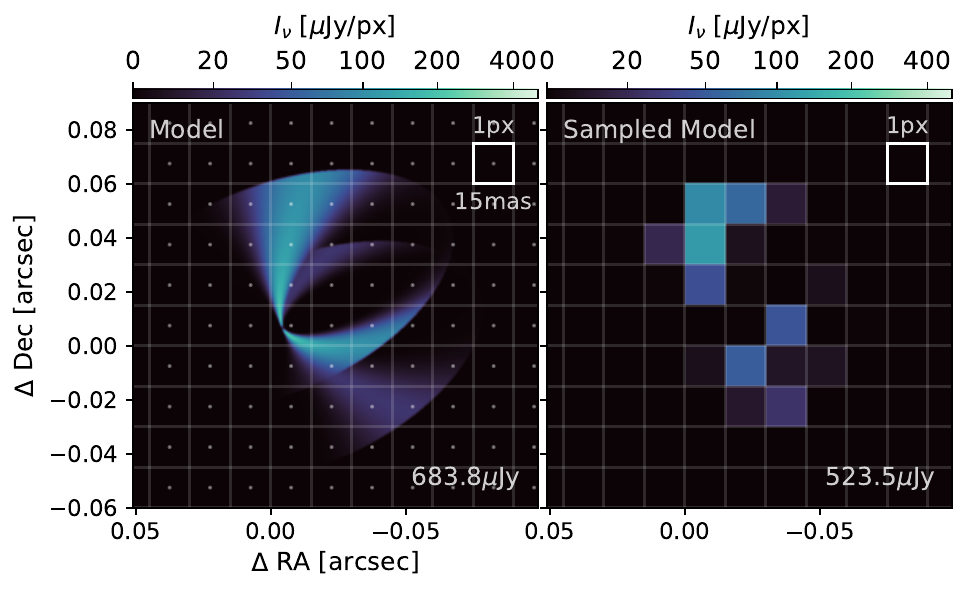}\\ \vspace{-0.1cm}
    \caption{Left panel: Image representative of the parametric model for the inner 10\,au intensity distribution of a single velocity channel of MHO\,6 (centered at 6\,km\,s$^{-1}$). The total flux from the image is shown in the lower right corner. The white lines represent the pixel grid used to sample the intensity distribution, prior to calculating its visibilities. The white dots show the location of the pixels center. 
    Right panel: Sampled model image, where each pixel has the value of the intensity distribution evaluated at the pixel center. Each pixel is 15\,mas per side, close to 2\,au at the distance of MHO\,6. Due to an incomplete intensity sampling, the total flux is underestimated.}
   \label{fig:sampling_ex}
\end{figure}

\begin{figure*}[t]
\centering
    \includegraphics[width=17.6cm]{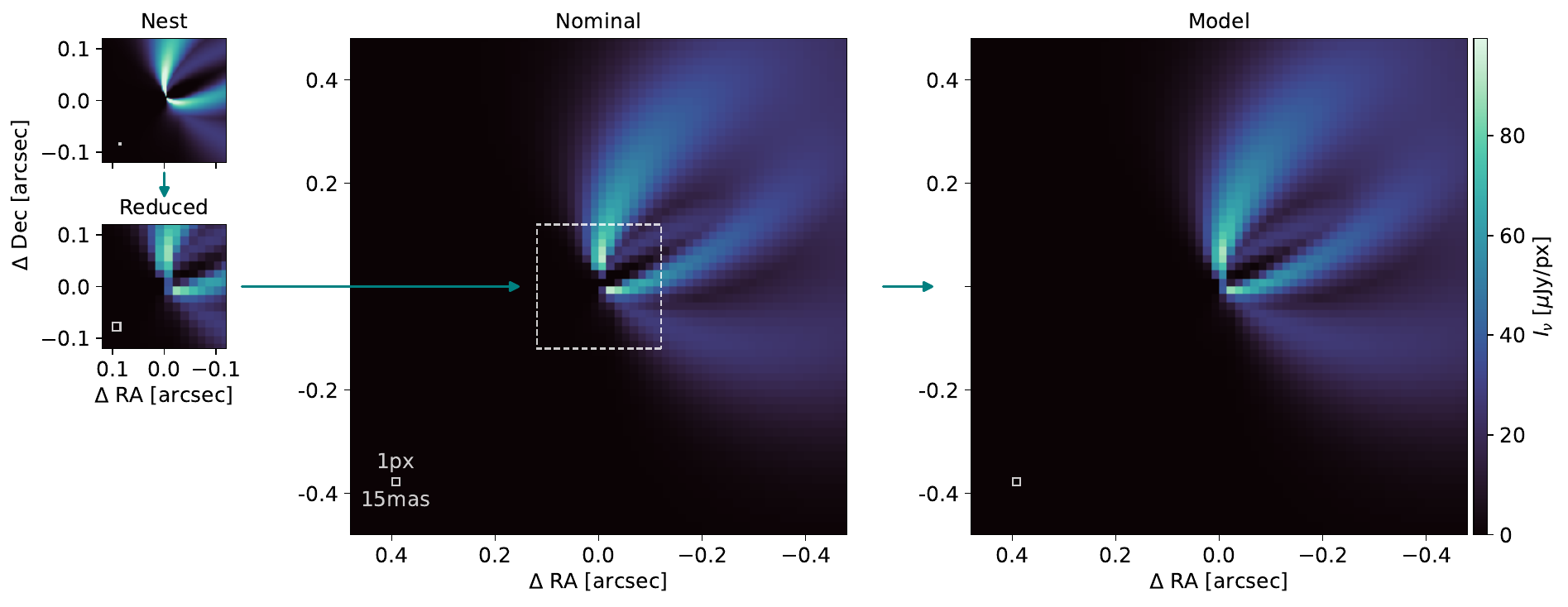}\\ \vspace{-0.1cm}
    \caption{Example of the nesting image process. The nest image is averaged into the reduced image, which then replaces the center of the nominal image to obtain the model image. In this example, the nest image was generated with a nest of level $n=3$. The pixel size is shown with a square in the bottom left corner of each panel. }
   \label{fig:diagram_nesting}
\end{figure*}

\subsection{Elevation, intensity, and combination of emitting layers}\label{sec:methodology_model}

The intensity distribution was calculated over a coordinate system that considers the elevation of the emitting layers. Those coordinates were calculated based on the functions from the python package \texttt{eddy} \citep{eddy}, which are organized for image cube generation in the \texttt{git-hub} repository \texttt{wrinkleology} \citep{simpledisk}. Two emitting layers are calculated for each line: front (upper) and back (bottom), one on each side of the midplane. The height of the emission surface $z(r)$ is described with an exponentially tapered power law as a function of the radial distance from the star $r$, as its commonly done in the literature \citep[e.g.,][]{eddy, Law2021, Izquierdo2021}. As an approximation for both CO isotopologues, we imposed the front and back layer should have the same height ($z_b=-z_f$) relative to the disk midplane (see Section \ref{sec:discussion} for a discussion on this decision). 
The equation describing $z_f$ is thus given by:
\begin{equation}
    z_f(r) \, = \, z'_{f0} \cdot \left( \frac{r}{r_0} \right)^{\psi} \cdot \exp\left( -\left( \frac{r}{r_z} \right)^{\phi} \right) \text{,}
    \label{eq:z_f}
\end{equation}
\noindent where $z'_{f0}$ is the elevation of the emitting layer at a distance $r_0$, before being tapered by the exponential part of the function.  
This function has a degeneracy between the parameters $z'_{f0}$ and $r_0^\psi$, which can be combined into $z_{f0}=z'_{f0}/r_0^\psi$, the equivalent to fixing $r_0=1''$ for a radial distance given in units of arcseconds \citep[as in][]{Law2021}. This reduces the number of free parameters describing the elevation of the emitting surface to four: $(z_{f0}, \psi, \phi, r_z)$. 

Next, we set the intensity of the front layer $I_f(r)$ of each pixel as a function of its central coordinates in the disk-frame. We chose to also describe it with an exponentially tapered power law, following: 
\begin{equation}
\centering
    I_f(r) \, = \, I'_{f0} \cdot \left( \frac{r}{r_0} \right)^{\alpha} \cdot \exp\left( -\left( \frac{r}{r_I} \right)^{\beta} \right) \text{.}
    \label{eq:I_f}
\end{equation}

During the calculation, the transformation, $I_{f0}=I'_{f0}/r_0^\alpha$, is also applied, with the same assumption over $r_0$ as in $z_f$. For the back layer, the intensity, $I_b$, is also described as in Eq.~\ref{eq:I_f}, but with different values for $(I_{f0}, \alpha, \beta)$ and sharing $r_I$ with $I_f$. 

To calculate the emission of individual channel maps, the intensity of the emitting layers ($I_f$, $I_b$) has to be filtered as a function of the projected velocity of each pixel $v_{proj}$. 
In the current model, the velocity field is approximated to be purely Keplerian, which is only consistent with disks where $M_{\text{disk}}\ll M_\star$, and without local perturbations from stellar companions or planets. 
Additionally, we assumed a Gaussian line-broadening in the velocity, where the linewidth of the emission in each pixel is described as a function of the radii:
\begin{equation}
\centering
    \delta v(r) \, = \, \delta v'_{0} \cdot \left( \frac{r}{r_0} \right)^{-q} \text{,}
\end{equation}
\noindent where $\delta v'_0$ is the linewidth at the distance $r_0$, which we fix to $r_0=1''$. 
Then, for a certain channel $k$ with a central velocity $v_k$, the contribution of a pixel with projected velocity $v_{proj}$ to the intensity of the front layer emission $I_f$ is given by:
\begin{equation}
\centering
    I_{f,k} \, = \, I_f \cdot \exp\left( - \frac{1}{2} \left( \frac{v_{proj} - v_k}{\delta v} \right)^2 \right) \text{,}
\end{equation}
\noindent where we assumed that $I_f$ has been projected onto the sky plane, and is described in the same coordinate system as $v_{proj}$. The same calculation is performed to obtain $I_{b,k}$, which describes the back layer emission. We include an additional parameter to describe a possible cavity in the the $I_f$ and $I_b$ emission, called $r_{cav}$. This parameters shifts the beginning of the brightness profile by a distance of $r_{cav}$ from the center. 

After obtaining the emission of each layer in a specific channel $k$, the next step for generating the model image of the channel is to combine the emission of the front and the back layer into a single image. 
For simplicity, we assume that the intensity value of each pixel in the model image is the maximum between upper and lower emission surface: 
\begin{equation}
\centering
    I_k = 
    \left\{
    \begin{aligned}
        I_{f,k}\,\text{,} & \hspace{0.6cm} \text{for } I_{f,k}\geq I_{b,k} \\
        I_{b,k}\,\text{,} & \hspace{0.6cm} \text{for } I_{f,k}<I_{b,k}\\
    \end{aligned}
    \right.\text{,}
    \label{eq:I_k}
\end{equation}
\noindent which is a first approximation for an optically thick front layer over the whole channel velocity width. 
This assumption is also considered in similar frameworks, such as \texttt{discminer} \citep{Izquierdo2021} and it has shown to be a good first approach to describe the $^{12}$CO emission from young circumstellar disks \citep[e.g.,][]{Izquierdo2022}. An example of combining layers in a single channel through this method is shown in the left panel of Fig.~\ref{fig:sampling_ex}. We also attempted an alternative combination, different from this binary selection between $I_{f,k}$ and $I_{b,k}$, with a smooth transition with a power law weighting. This alternative did not produce significant improvements over the fitting, and its details are described in the Appendix \ref{app:sec:appendix_coord}.

As the CO lines from MHO\,6 are affected by surrounding cloud absorption, this contribution needs to be considered in order to fit the flux distribution of each channel. We assume that the absorption decreases the emission of the whole channel by a proportional factor as a function of velocity, which we describe with a Gaussian function ($G_{ext}$). This parametrization of the absorption has three free parameters: the velocity center ($c_{v0}$), the velocity width ($c_\sigma$), and the amplitude of the absorption ($c_a$), which we fit for both CO isotopologues. 
No cloud contribution in positive emission was considered in our fitting.

\begin{figure*}[t]
\centering
    \includegraphics[width=18cm]{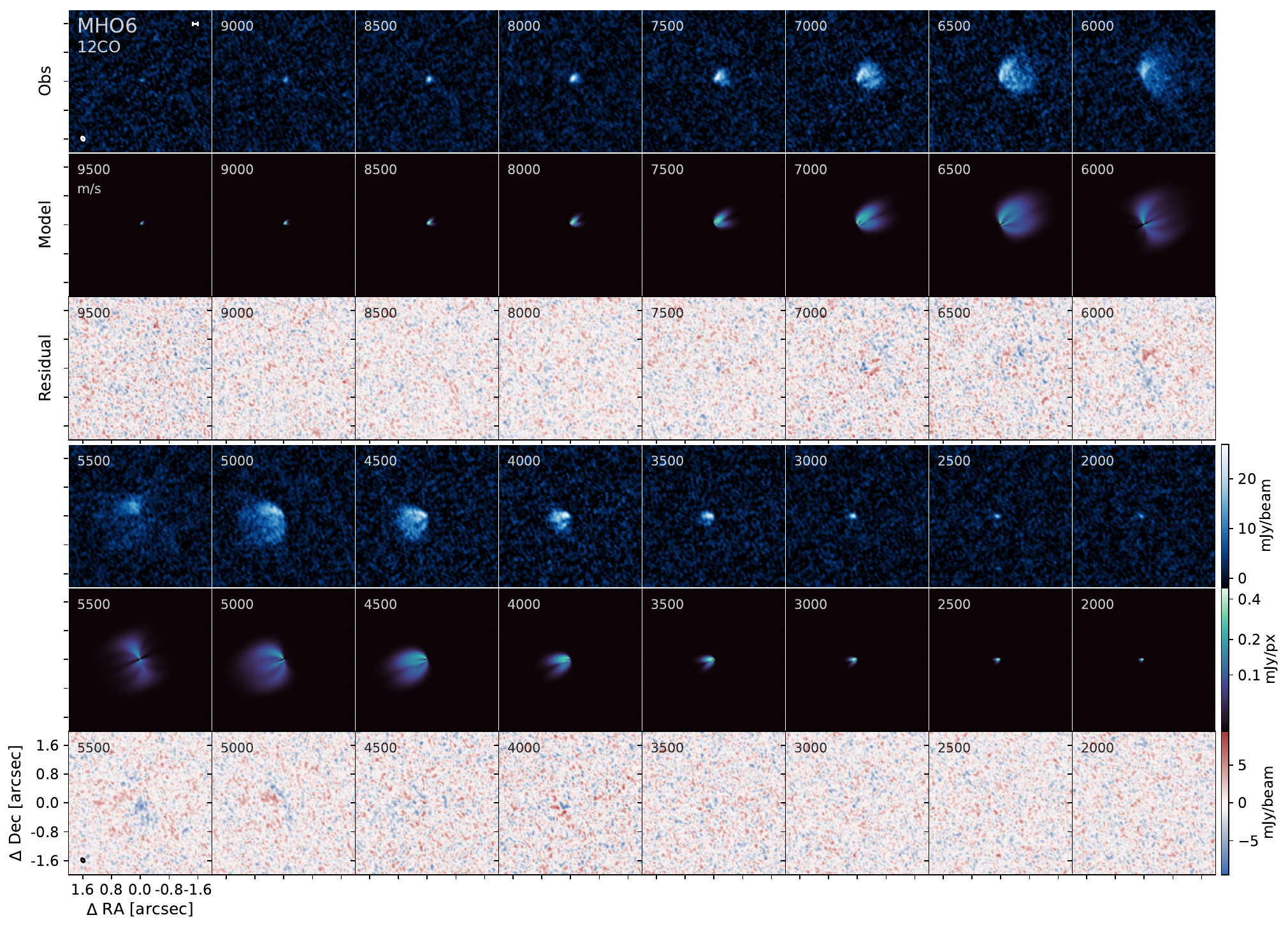}\\ \vspace{-0.1cm}
    \includegraphics[width=18cm]{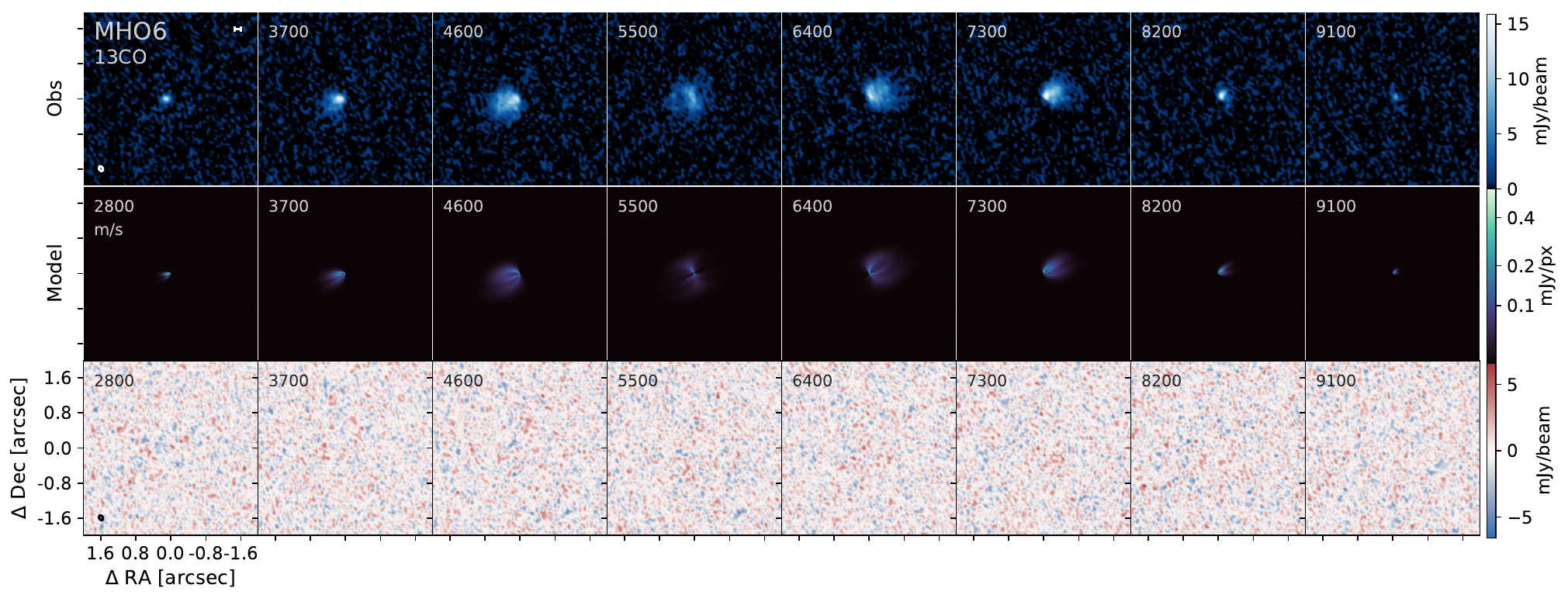}\\ \vspace{-0.1cm}
    \caption{ $^{12}$CO J=3-2 and $^{13}$CO  J=3-2  channel maps. Each row shows the CLEAN image of the observation, the best visibility model, and the image made from the residual visibilities. The scale bar in the upper left panels is 10\,au in size, while the beam size is shown in the lower left corner of the first panels. The number in the upper left corner shows the channel velocity relative to the resting frequency of each line, in m\,s$^{-1}$.}
   \label{fig:chanmap}
\end{figure*}

\begin{figure*}[t]
\centering
    \includegraphics[width=18cm]{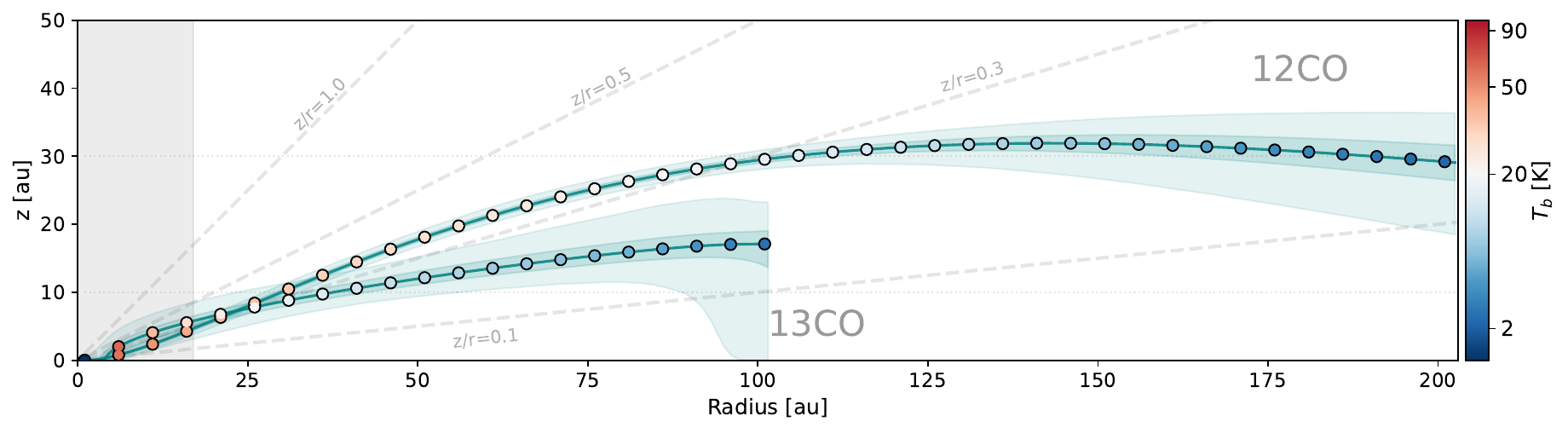}\\ \vspace{-0.1cm}
    \caption{ Height of the surface emitting layer for the $^{12}$CO and $^{13}$CO of MHO\,6. 
    The  darker and lighter shaded regions surrounding each profile represent the $1\sigma$ and $3\sigma$ confidence levels, respectively. The grey shaded region on the left shows the spatial extent of one angular resolution element, as measured from the disk center. The distance in the x-y axis is shown in a 1:1 scale.}
   \label{fig:height_COs}
\end{figure*}

\subsection{Nesting images}\label{sec:nesting_images}

The method of visibility fitting is based on matching the visibilities of an observation to those calculated from a given intensity distribution model. Since the model visibilities are calculated numerically, an underlying assumption is that the brightness associated with each pixel is representative of the flux contained inside  the region covered by the pixel extent. As an example, let us assume a parametric function that describes $I_k$ (the intensity of a channel $k$), represented in the left panel of Fig.~\ref{fig:sampling_ex} with extremely high spatial resolution. In order to numerically calculate the visibilities of this channel, we need to create an image of $I_k$ (the ``sampled Model''), shown in the right panel of Fig.~\ref{fig:sampling_ex}. The pixel size of 15\,mas is comparable to the pixel size in high angular resolution observations of gas tracers with ALMA. In the sampled model, the value of each pixel is the value of the function at the center of the pixel; thus, if the pixel size is larger than the scale of spatial variations in the intensity model, the flux contained in each pixel will not be representative of the integrated flux inside such a region. This effect commonly leads to flux underestimation in the inner regions of the disk, as shown with a 10\,au disk in Fig.~\ref{fig:sampling_ex}, and it can influence the fitting even if the affected region is spatially unresolved. To overcome this problem, the pixel size needs to be small enough to sample each channel intensity distribution. In a single image, such pixel size can be prohibitively small for the inner disk region, forcing the  number of pixels contained in an image to be unnecessarily large to properly cover the outer regions of the disk. 

We explored a solution to the issue of Fig.~\ref{fig:sampling_ex} by generating two images of the same intensity distribution: The first covers an extended field of view that includes the whole disk emission, while the second is smaller and has a higher pixel resolution to sample the inner disk region. In the following, we refer to the first image as the ``nominal'', and the second as the ``nest,'' as exemplified in Fig.~\ref{fig:diagram_nesting}. To combine both, we averaged the pixels of the nest image to match the pixel size of the nominal, creating a ``reduced'' image. Then, the central section of the nominal image is replaced with the reduced image, obtaining a ``model'' image with the same image size and pixel size as the nominal, but with a higher flux fidelity. We further refer to this process as ``nesting'' images. 

We explored oversampling the pixels of the nominal image in multiples of 4. Thus, an image with no nesting would be the same as oversampling with a factor of $4^0$; one level of nesting would mean that every pixel in the nominal image is oversampled with $4^1$ pixels; and, more generally, an $n$ level of nesting means that every nominal pixel in the central region is oversampled with $4^n$ pixels. An example of the difference between using different levels of nesting, up to level 6, is shown in Fig.~\ref{fig:app:nested_examples} in the appendix, where the flux of each pixel in the reduced image is compared to that of an image generated with model with a nesting level of 6. A more detailed exploration of flux conservation with nested images is also given in Section \ref{app:sec:nesting_images} in the Appendix. For MHO\,6, we used a nominal image of 512\,px in size, with each pixel being 15\,mas. The inner 64 pixels of the nominal image are replaced by a nest image of level $n=2$, which means that it oversamples the nominal pixel with $4^{n=2}$ pixels. The size of the region covered by the nest image is determined empirically, and we discuss this topic further in Sect.~\ref{sec:discussion:how_much_nesting}.

\subsection{MCMC fitting}\label{sec:methodology_mcmc}

We fitted both CO isotopologues independently, with the same parametric functions. Thus, parameters such as the disk geometry ($\Delta$RA, $\Delta$Dec, inc, PA), stellar mass ($M_\star$), central disk velocity, and cloud velocity (VLSR, c$_{v0}$), are allowed to be different for each line, to test for consistency. We sampled each parameter density distribution with the Markov chain Monte Carlo (MCMC) code \texttt{emcee} \citep{emcee2013}, using eight times the number of walkers with respect to the free parameters. Each parameter has a flat uniform prior over the allowed parameter range, which extends far enough so that the walkers do not interact with it. Each model results are sampled from the last $10^5$ steps, after a burning period of at least $10^5$ steps.


\section{Results \label{sec:results}}

The results of our MCMC fitting are shown in Fig.~\ref{fig:chanmap} for a subset $^{12}$CO and $^{13}$CO channels around the central velocity. Both CO isotopologues find consistent solutions for the shared parameters, such as the midplane geometry, the stellar mass, $M_\star$, and the central velocity, VLSR. The $^{12}$CO emission constrains  $M_\star=0.164\pm0.003$, which is consistent with the $^{13}$CO modeling result, and also with the literature \citep{wardduong2018, kurtovic2021, pegues2021}. Furthermore, the models successfully disentangle the regions corresponding to the front and back emitting layers, which allows us to estimate their brightness temperature profiles. The front emitting layer of the $^{12}$CO emission reaches its maximum height at a radii of about $100$ to $125$\,au, close to the $z/r=0.3$ ratio, after which it starts declining towards the midplane, as shown in Fig.~\ref{fig:height_COs}. Starting from about 20\,au, the best model for the $^{13}$CO emission has a height below that of the $^{12}$CO. Within the first 20\,au, which closely coincides with the spatial resolution of the dataset, the height of the $^{12}$CO and $^{13}$CO are consistent within the $1\sigma$ limit. We note that the turnover parameters $\beta_f$ and $\phi$ of the $^{13}$CO emission are very unconstrained, as the turnover distances, $r_z$ and $r_I$, coincide with the $R_{95\%}$.

Our fitting also recovers the magnitude of cloud transparency as a function of velocity. As expected, the $^{13}$CO emission is less affected by the cloud absorption, with the lowest transparency being approximately $70\%$ at the central velocity of the disk. On the other hand, the $^{12}$CO emission has the lowest transparency of $50\%$ at the central velocity channel. By knowing the magnitude of the absorption, we can correct the measured fluxes of each disk, as listed in Table~\ref{tab:mcmc_results}. All our analyses of the disk sizes and temperature were carried out with the absorption-corrected models.

\begin{figure}[t]
\centering
    \includegraphics[width=9cm]{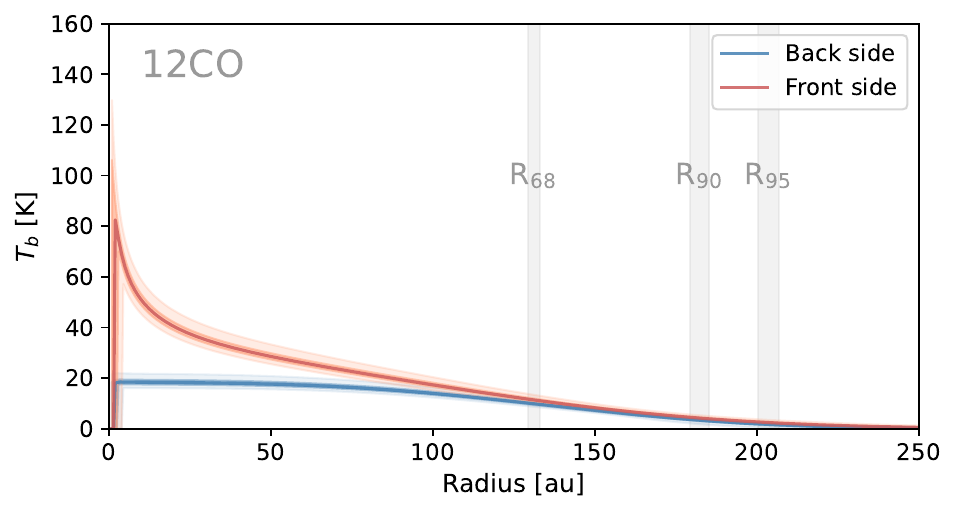}\\ \vspace{-0.1cm}
    \includegraphics[width=9cm]{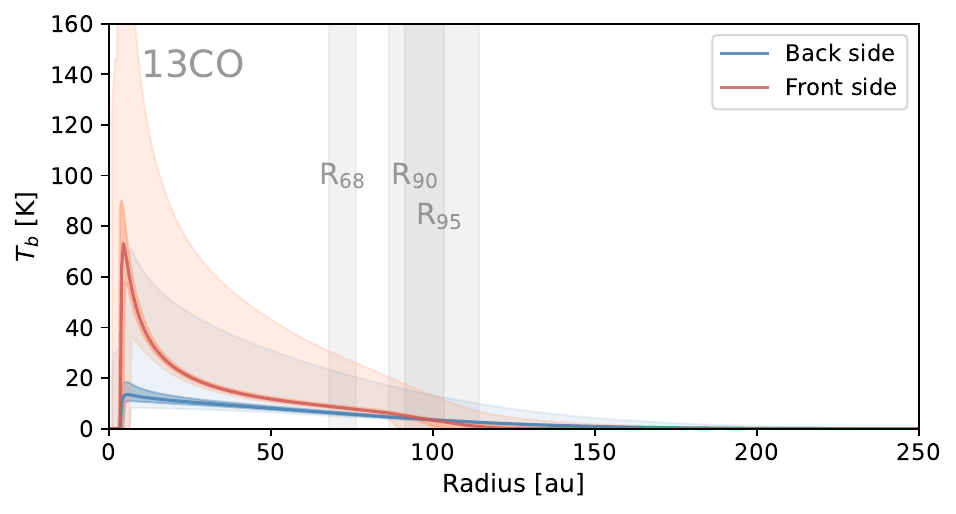}\\ \vspace{-0.1cm}
    \caption{  Azimuthally averaged   brightness temperature profile for the $^{12}$CO and $^{13}$CO emission in the upper and lower panel, respectively. The  darker and lighter shaded regions represent the  $1\sigma$ and $3\sigma$ uncertainty of the brightness temperature, respectively. The solid line shows the best model. The vertical bands in grey show the $3\sigma$ radii region that includes the 68\%, 90\% and 95\% of the emission.}
   \label{fig:profiles}
\end{figure}

Our channel-by-channel model describes the $^{13}$CO emission down to the sensitivity level of the observation, leaving no significant residuals in the reconstructed image. In the $^{12}$CO emission, however, structured low-contrast non-axisymmetric residuals are detected around the central velocity channels, which we further discuss in Sect.~\ref{sec:discussion}. 
We calculated the brightness temperature of each model by converting the emission from intensity units in Jy\,px$^{-1}$ to temperature in Kelvin. This was done under the Rayleigh-Jeans approximation for black body emission in the low frequency regime, assuming the emission is entirely optically thick. Thus, the azimuthal profile of the brightness temperature is directly recovered from our fitting to both CO isotopologues, as shown in Fig.~\ref{fig:profiles}.

The size of the disk is measured by constraining the radii enclosing $68\%$, $90\%$, and $95\%$ of the total flux as measured from the front layer emission surface. Those values are listed in Table~\ref{tab:mcmc_results} for both lines and also shown to their $1\sigma$ confidence in Fig.~\ref{fig:profiles}. Despite using a different definition when measuring the disk size, we found values that are consistent with those measured from integrating the moment 0 emission with an elliptical aperture in \citet{kurtovic2021}. 
Therefore, we confirm the gas to dust size ratio of 3.9 when comparing the 0.87\,mm emission to the $^{12}$CO J=3-2 emission. It is not uncommon to observe such a high ratio among young disks \citep{long2022} and it suggests that radial drift has influenced the disk evolution \citep{trapman2019}. Interestingly, both the front and back emission layer converge to the same brightness temperature at the $R_{68\%}$ size of the front layer in both CO isotopologues.


\section{Discussion \label{sec:discussion}}

\subsection{Morphology of MHO6 CO isotopologues emission}

The elevation and intensity of the CO isotopologues emitting layer is well described by an exponentially tapered power law in radius. The $^{13}$CO exhibits no significant residuals after model subtraction, and only low-contrast residuals are observed in the $^{12}$CO channel maps. There is a consistent difference in the brightness temperature distribution of the front and back layers in both CO isotopologues, as illustrated in Fig.~\ref{fig:profiles}. Notably, the brightness temperature of the back layer maintains a nearly constant value of $18\pm1$\,K within the first 100\,au of $^{12}$CO emission. 
This temperature aligns with the expectations for an optically thick layer tracing the region where CO freezes onto dust grains, and is consistent with the 17\,K detected in TW\,Hya \citep{qi2013} and 21\,K observed in IM\,Lup \citep{pinte2018}. This unprecedented measurement from a moderate resolution and moderate sensitivity observation suggests a cold midplane with $T<20$\,K. A similar nearly constant value is observed in $^{13}$CO emission within the initial 50\,au, followed by a rapid decline, likely due to optical depth effects.  We note that our model does not consider possible absorption in the midplane to the emission from the back layer, which could be lowering the observed brightness temperature.

Our model for the brightness temperature becomes highly uncertain for $r<10$\,au in both CO isotopologues, due to the limited angular resolution and fitted velocity range. At all radii, the brightness temperature of the $^{12}$CO front layer is consistently higher than that of the $^{13}$CO front layer (see Fig.~\ref{fig:profiles}), as well as more elevated from the midplane (see Fig.~\ref{fig:height_COs}), therefore suggesting the detection of a vertical temperature gradient \citep[as in ][]{Law2021, paneque2023}. It is unclear, however, how the optical depth contributes to this result. Additional observations with alternative gas tracers could help disentangle this degeneracy and provide a robust constraint on the vertical temperature structure of the disk.  Additionally, our parameter $r_{cav}$ does not find a robust detection of an empty cavity, as it is consistent with being $r_{cav}=0$ within the $3\sigma$ given by the MCMC.

In the dust continuum emission, MHO\,6 shows a ring peaked at 11\,au and a resolved cavity \citep{kurtovic2021}. Our visibility models of the CO isotopologues, however, do not find robust conclusive evidence of an emission cavity, at least at the sensitivity and resolution of the observations. One potential explanation could involve a combination of high optical depth in the inner regions and the presence of gas flowing from the millimeter dust ring location to the cavity. A higher angular resolution observation is required to test this hypothesis.

The residuals of our models do not show evidence of additional ringed structures, as observed in the $^{12}$CO emission of other Class II disks \citep[e.g.,][]{teague2019, Law2022a}. However, we do observe non-azimuthally symmetric structured residuals after subtracting the best model of $^{12}$CO emission. These residuals form an ``X'' pattern when collapsed in a moment 0 image, as shown in Fig.~\ref{fig:app:momres} in the appendix. 
This pattern is a known radiative transfer effect generated by optically thick emission originating from two emitting layers, and it is commonly observed in CO isotopologues moment 0 images of disks with mid-inclinations \citep[see ][]{keppler2019, law2021a, zhang2021}. Detecting this ``X'' residual suggests an incomplete description of the interaction between emission from the front and back layer, which we combined using the binary selection shown in Eq.~\ref{eq:I_k}.

An additional experiment was attempted to combine both layers using only two additional free parameters, assuming a simple relation between optical depth and intensity, as explained in Appendix~\ref{app:sec:appendix_coord}. 
However, this model did not significantly improve the residual structure compared to Eq.~\ref{eq:I_k}, suggesting that a more detailed approach might be necessary, including independent optical depth variations in radii and azimuth. 
Due to its low brightness contrast and non-azimuthal structure, the ``X'' shape is unlikely to significantly influence the results obtained for $h(r)$ and $I(r)$. Nonetheless, its detection after subtracting the best model demonstrates the sensitivity of visibility-based approaches to very subtle radiative transfer effects. This opens the possibility of conducting temperature and optical depth studies even in moderate sensitivity observations.

\begin{figure*}[t]
\centering
    \includegraphics[width=18cm]{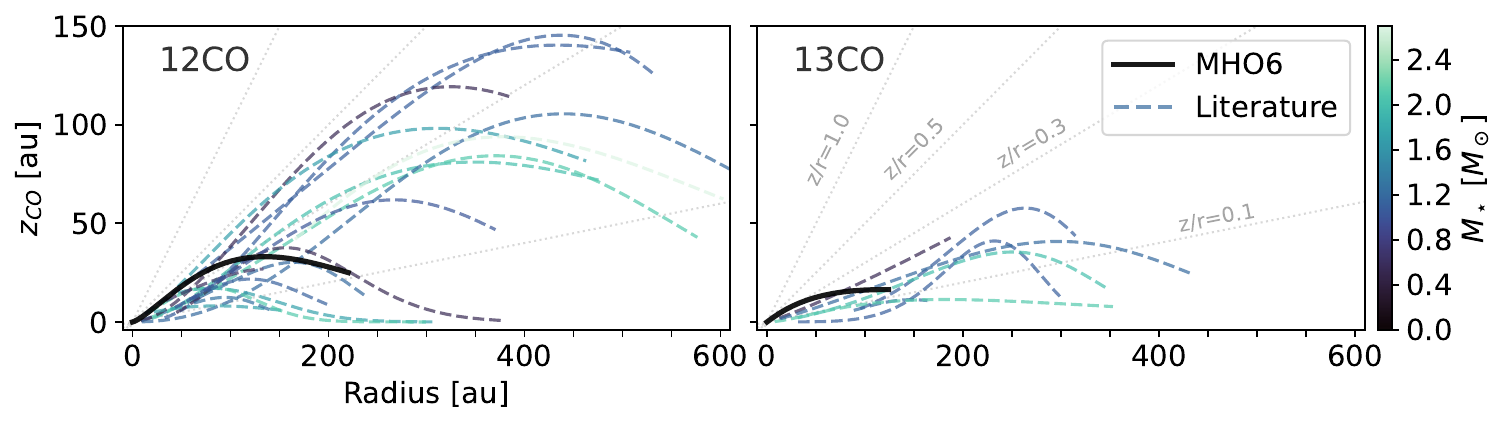}\\ \vspace{-0.1cm}
    \caption{ Line emission surface height of the MHO\,6 disk compared to the fitted surfaces of $^{12}$CO and $^{13}$CO emission from \citet{Law2021, Law2022a, Law2023}. Surfaces are only shown over their fitting range, while MHO\,6 is shown until its $R_{95\%}$. The color of each line represents the mass of the central star. }
   \label{fig:surf_height}
\end{figure*}

\begin{figure*}[t]
\centering
    \includegraphics[width=15cm]{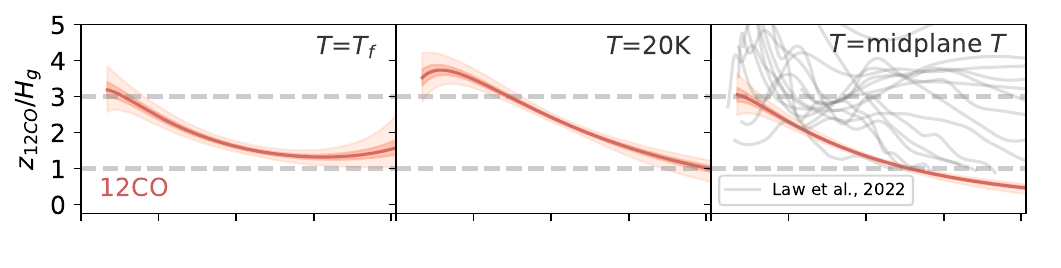}\vspace{-0.7cm}
    \includegraphics[width=15cm]{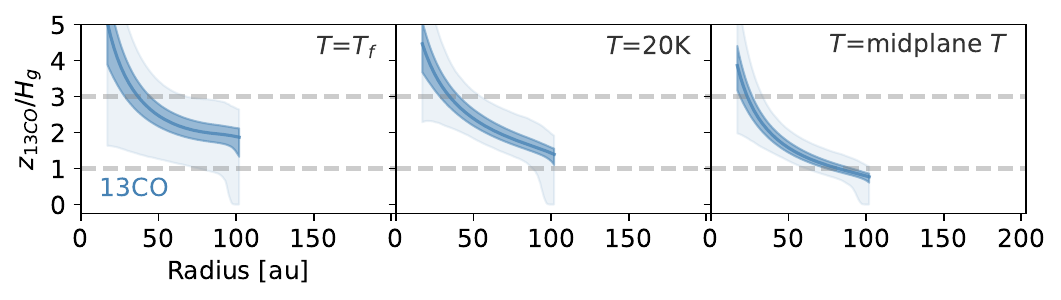}\vspace{-0.1cm}
    \caption{Ratio between the elevation of the emitting layer $z(r)$ over the expected pressure scale height $H_g$ for different midplane temperature prescriptions. Left: Temperature profile equal to the front layer temperature of each isotopologue.  Middle: Temperature constant equal to 20\,K. Right: Temperature follows Eq.~\ref{eq:tmid}. The darker and lighter shaded regions represent the $1\sigma$ and $3\sigma$ uncertainty of each profile, respectively. The gray lines in the top right panel show the $z/H$ ratio from the disks analyzed in \citet{Law2022a}. }
   \label{fig:hg}
\end{figure*}

\subsection{Comparison of MHO6 with other sources}

MHO\,6 is the first very low mass star for which there is a direct measurement of the elevation of its emitting layer and brightness temperature as a function of its radii. In the following, we compare some of its measured properties to other disks that have been observed with ALMA. 

The elevation of the emitting surface $z(r)$ of MHO\,6 was modeled with the same functional form as in \citet{Law2022a, Law2023} in the image plane, which allows for a direct comparison of the results. The disks studied in those works are Class II sources with mid-inclination midplanes relative to the line of sight, covering a range of stellar masses from $0.5\,M_\odot$ to $2.7\,M_\odot$. 
In Fig.~\ref{fig:surf_height}, we compare the elevation of the emitting surfaces of those disks to our measurement in MHO\,6, which is the star with the lowest mass of the sample ($0.164\,M_\odot$ from $^{12}$CO). Within the first 50\,au in radii, MHO\,6 is among the most elevated emitting surfaces in both CO isotopologues when measured in distance from the midplane. This result is consistent with models which predicted disks around VLMS to be geometrically thicker due to the reduced force of gravity from the central star \citep{mulders2012, sinclair2020}. 
The $^{12}$CO emitting layer is at $z/r>0.3$ until a radii of 100\,au, where it starts decreasing due to the more compact disk extension compared to the other sources. 

Next, we investigated the ratio between the height of the emmitting surface and the gas pressure scale height $H_g$. 
As in \citet{Law2022a}, we describe $H_g$ as a function of radii under the assumption of vertical hydrostatic equilibrium, following: 
\begin{equation}
    H_g = \sqrt{\frac{k_B\,T_{\rm{mid}}\,r^3}{\mu\,m_p\,G\,M_\star}} \text{,}
\label{eq:Hg}
\end{equation}

\noindent where $k_B$ is the Boltzmann constant, $G$ the gravitational constant, $m_p$ is the mass of the proton, and $\mu$ the mean molecular weight of the studied species. The only unknown quantity is the value of $T_{\rm{mid}}$ as a function of radii; this is because the mass of the star, $M_\star$, is constrained by our model. For each CO isotopologue, we calculate three different $H_g$ as a function of radii, by considering different descriptions for the midplane temperature. First, we considered the temperature to be that of the isotopologue front layer, as an approximation of a vertically isothermal disk; second, we set the temperature to be constant equal to 20\,K; and third, we assume the midplane temperature is described by the simplified expression for a passively heated, flared disk in radiative
equilibrium \citep[][]{kenyon1987, chiang1997, dalessio1998, dullemond2001}, as in \citet{Law2022a}: 
\begin{equation}
    T_{\rm{mid}} \,=\, \left( \frac{\varphi L_\star}{8 \pi r^2 \sigma_{SB}} \right)^{1/4} \text{,}
    \label{eq:tmid}
\end{equation}

\noindent where $\varphi$ is the flaring angle, which we set to 0.02 for comparison with the literature, $\sigma_{SB}$ is the Stefan-Boltzmann constant, and $L_\star$ is the stellar luminosity. By using $L_\star=0.06L_\odot$ from \citet{kurtovic2021}, derived from \citet{Herczeg2014}, Eq.~\ref{eq:tmid} returns a $T_{\text{mid}}$ ranging between $20$  to  $5$\,K between the radii of $10$ to $150$\,au, being the lowest of the three temperature descriptions. 

Despite having one of the most elevated emission surfaces, MHO\,6 shows one of the lowest ratios of $z/H_g$ among the sample of disks where this ratio has been measured, as shown in Fig.~\ref{fig:hg}. For the $^{12}$CO emission, the main source of uncertainty to estimate the scale height of the emission layer comes from the $T_{\rm{mid}}$ assumption. Depending on the temperature prescription, we find that $z/H_g$ ranges from $1$ to $3\times H_g$. This estimate of an average $z/H_g\approx2$ is consistent with the relations found in \citet{Law2022a}, where $z/H_g$ changes weakly with $M_\star$, and has a positive correlation with $R_{90\%}$. 
In the case of $^{13}$CO, the ratio $z/H_g$ is poorly constrained due to the uncertainty in $z_{13CO}(r)$, although the estimates appear to prefer the region of $1$ to $4\times H_g$ to a $1\,\sigma$ of uncertainty. 

As expected for a very low mass star, MHO\,6 shows a very low $^{12}$CO brightness temperature when compared to its more massive counterparts. At 100\,au, the brightness temperature of the $^{12}$CO emission is $17\pm1$\,K, a few degrees lower than the coldest disk from the \citet{Law2022a} sample (GW\,Lup, with $26\pm0.5$\,K at 100\,au). However, the role of the optical depth in this measurement is still unclear, given that at 100\,au the brightness temperature of the front and back layer becomes comparable (see Fig.~\ref{fig:profiles}).

\subsection{Advantages of image- versus visibility-based analysis}

Among the advantages of analyzing gas emission through parametric visibility modeling, the most notable is the incorporation of prior information on the model morphology. 
Typical image reconstruction methods, such as CLEAN, generate individual models for each channel without including information on the brightness distribution in the surrounding velocity channels. Consequently, the model of each channel is constrained by the individual channel sensitivity. This conservative approach is advantageous when the brightness distribution does not follow a characterizable spatially coherent structure in space and velocity, as observed in disks with strong Keplerian deviations, inflows, or outflows. However, it also limits the recovery of disk-like emission structures that are spread over many velocity channels. The generation of additional imaging products, such as moment maps, tends to limit the sensitivity of those collapsed images to be, at best, that of a single channel due to the cumulative effect of adding thermal noise. In contrast, prior assumptions over the emission morphology enable the characterization of faint structures that span several channels, substantially increasing the detection sensitivity.

The advantages that come with the implementation of a disk-like prior for the emission morphology can also be obtained through parametric image modeling. Previous works have used this approach to recover super-resolution structures in the inner disk from observations with high frequency resolution \citep[e.g.,][]{bosman2021}, or to estimate deviations from expected pure Keplerian motion in the outer disk \citep[e.g.,][]{Izquierdo2021}. The difference between parametric image modeling and parametric visibility modeling comes in the stage at which the disk-like prior is assumed. In the image-based analysis, the prior is assumed after generating a model image for the visibilities of each channel. In a visibility-based analysis, the prior is assumed when generating the model image, thus preventing image reconstruction issues from influencing the model properties. Most of those issues are related to the beam convolution of CLEAN: (i) The beam convolution can artificially erase resolved emission structures that are well detected in the visibility plane \citep[e.g.,][]{andrews2021, jennings2022}; (ii) an inclined beam can modify the brightness distribution by enhancing the emission over the beam's major axis; (iii) the emission can dilute within the beam if the beam area is larger than the emitting area, such as in the inner regions of the disk, leading to an underestimation of the brightness temperature; and (iv) faint large scale emission structures can be diluted in the noise background due to small beam sizes, even if the emission is detected in the visibilities \citep[e.g., the outer ring of TW\,Hya,][]{ilee2022}. Compared to the image plane, fitting in the visibility plane has the advantage of fitting uncorrelated noise.  
Although the inclined beam problem can be mitigated by forcing circularization of the point spread function \citep[e.g., with visibility tappering, as in MAPS, ][]{czekala2021}, this circularization only increases the image spatial resolution loss. As for the dilution of emission in small and large scales, a typical imaging approach is to increase the weight towards more compact or extended baselines or to exclude specific ranges of the sampled visibilities. The visibilities, however, have the additive properties of the Fourier transform, and thus, excluding visibilities or modifying the weight given to them during the imaging process leads to further loss of total information. By creating and fitting a model in the visibility plane, the model must simultaneously fit all the visibility coverage, which mitigates issues related to spatial information loss and beam dilution. At the same time, the brightness distribution is controlled from the parametric model, avoiding biases from a non-symmetric visibility coverage. 

Fitting all spatial scales covered by the visibilities with a disk morphology also has limitations. In an image-based approach, certain complex disk regions can be masked out of the fit, commonly isolating the outer or inner disk emission \citep[e.g.,][respectively]{teague2021, bosman2021}, thus simplifying the models to focus them on a particular science case. Masking spatial regions is not possible with visibility modeling. 
As an example of adding complexity to a model, our visibility fitting of MHO\,6 needed to include cloud contamination to the line emission as additional free parameters. If not accounted for, the model fitting would fit the cloud contribution by changing other disk parameters, such as the inclination or elevation of the emitting surface, thus misleading the recovery of disk information.

Another relevant comparative aspect of image versus visibility-based parametric modeling is their scalability with data volume. For a fixed frequency resolution, the fitted data volume in an image is determined by the image size and pixel size, as increasing the exposure time of the observation will result in higher image sensitivity, not in larger images. Further data volume reduction can be done by calculating the moment maps of the channels, which also allows for independent studies focused solely on the emission morphology or the kinematics. In the case of visibility analysis, increasing exposure time will continuously increase the fitted data volume due to the additive properties of the Fourier transform, and there is no feasible visibility simile to creating a moment map with the visibilities. Thus, a visibility model must describe intensity distribution and gas kinematics simultaneously. The selection of one method over another becomes dependent on the scientific goal. Overall, the advantages of visibility-based approaches are maximized when studying observations inaccessible to image-based techniques. One example is the recovery of the $^{13}$CO emission morphology of MHO\,6 shown in Fig.~\ref{fig:chanmap}, which despite being detected in several channels, does not allow for an image characterization of its structure due to the low S/N in each beam. Thus, emission detected with low sensitivity, or disk regions resolved by only a few resolution elements are the ideal study cases to be considered for visibility modeling.

\subsection{How much nesting is needed}\label{sec:discussion:how_much_nesting}

Increasing the level of nesting results in better pixel flux conservation in the inner disk, as exemplified in Fig.~\ref{fig:app:violin_diff}, where the flux ratio between a nested image of $n=6$ is compared to models with fewer levels of nesting. 
However, additional nesting levels come with additional computational cost. The acceptable flux loss should be decided on the basis of a comparison with an observation's sensitivity, such that the dominant uncertainty source is the single channel sensitivity and not the model's flux loss. 

The level of nesting needed to model a specific observation will be dependent on the spatial resolution, frequency resolution, and disk properties. The main driver for flux loss during the construction of a parametric model is the undersampling of emitting regions in the inner disk, when the scale of brightness variations becomes smaller than the pixel size. 
This effect should be most significant for observations with very high-frequency resolution towards disks with high midplane inclinations or small line broadening in the line of sight. 

In the case of the MHO\,6, a pixel size of 15\,mas is small enough to model the spatial frequencies covered in the visibilities, as the pixel is about seven times smaller than the angular resolution. However, as shown in Fig.~\ref{fig:app:nest_diff} and Fig.~\ref{fig:app:violin_diff}, the inner 10\,au can have a flux loss as large as 20\% with this pixel size, which is about 0.2\,Jy for the $^{12}$CO emission. 
Increasing the nesting level from $n=0$ to $n=2$ leads to a decrease in flux loss from 20\% to 2\%, which is about 0.022\,Jy for the inner 10\,au. By dividing over the 32 fitted channels, the nesting of $n=2$ leads to an average flux loss of $\sim0.7$\,mJy per channel, or $\sim1.8$\,K in brightness temperature considering that a region of 10\,au is contained in a single beam. Thus, the flux loss is below $0.5\sigma$ per channel, and is below the profile noise level in the inner 10\,au (see uncertainty in Fig.~\ref{fig:profiles}). Even though a nesting of $n=2$ is appropriate for this observation of MHO\,6, the value of $n$ should be considered for every source and every observation separately. 

An important consideration to determine $n$ is that the flux loss due to pixel undersampling is proportional to the total flux. Therefore, disks with very bright emission lines and high sensitivity observations could require a higher level of nesting compared to fainter disks observed at the same angular resolution, frequency resolution, and exposure time.


\section{Conclusions}
\label{sec:conclusions}

We present a novel method that uses parametric visibility modeling to analyze the emission profile of molecular lines obtained with interferometric millimeter observations of protoplanetary disks. To reach flux conservation with a non-prohibitive pixel size, we explored oversampling the central pixels using ``nested images,'' thereby increasing the flux fidelity of the models without using Monte Carlo-based radiative transfer methods. By remaining purely parametric, the modeling becomes rapidly iterable, which is key for testing different parametric combinations to describe the emission properties. 

We applied this new method to the observations of $^{12}$CO and $^{13}$CO of the protoplanetary disk around the very-low-mass star MHO\,6. Our method confirms the stellar mass of MHO\,6 obtained in previous works \citep{kurtovic2021, pegues2021}, independently obtaining consistent results with each isotopologue. When compared to the sample of \citet{Law2022a}, we find the surface of MHO\,6 to be among the most elevated surfaces in distance from the midplane within the first $50\,$au in radii. This is in agreement with the geometrically thicker morphology that is expected for disks around very-low-mass stars compared to Solar-type disks. We also recover the brightness temperature of the back emission layer in the $^{12}$CO emission, which is consistent with the freezing temperature of this molecule. 

Using the emission of our models, we obtain the gas disk radii and confirm that the ratio of gas to dust size is around 4, which has been suggested as evidence of dust evolution through radial drift \citep{trapman2019}. The emission in the $^{12}$CO shows evidence of a low-contrast non-azimuthally symmetric temperature distribution, which could be a result of an incomplete description of an optical depth effect. 

The robustness of a parametric visibility modeling approach to analyzing the gas emission of protoplanetary disks is demonstrated with our fit to the $^{13}$CO emission. Despite the $^{13}$CO being compact in size, and only detected in a few channels with poor frequency resolution, our model recovers consistent results compared to the $^{12}$CO model, providing precise measurements and uncertainties for the emission morphology, size, and flux. 
Visibility modeling represents an alternative approach to analyzing those observations where image-based methods are not able to disentangle emission surfaces, directly recovering all the relevant properties of the disk for comparison with models of planet formation.

\section*{Acknowledgments}

We thank the anonymous referee for the many useful comments. 
We also thank Richard Teague for his work on the packages \texttt{eddy} and \texttt{wrinkeology}, and for making them available for public use. 
N.K. and P.P. acknowledges support provided by the Alexander von Humboldt Foundation in the framework of the Sofja Kovalevskaja Award endowed by the Federal Ministry of Education and Research. 
This paper makes use of the following ALMA data: ADS/JAO.ALMA\#2018.1.00310.S. ALMA is a partnership of ESO (representing its member states), NSF (USA) and NINS (Japan), together with NRC (Canada), MOST and ASIAA (Taiwan), and KASI (Republic of Korea), in cooperation with the Republic of Chile. The Joint ALMA Observatory is operated by ESO, AUI/NRAO and NAOJ. P.P. acknowledges funding from the UK Research and Innovation (UKRI) under the UK government’s Horizon Europe funding guarantee from ERC (under grant agreement No 101076489).

\bibliographystyle{aa.bst}
\bibliography{ms.bib}


\onecolumn

\begin{appendix}

\section{Parametric visibility modeling results}

\begin{figure}
 \centering
        \includegraphics[width=9.5cm]{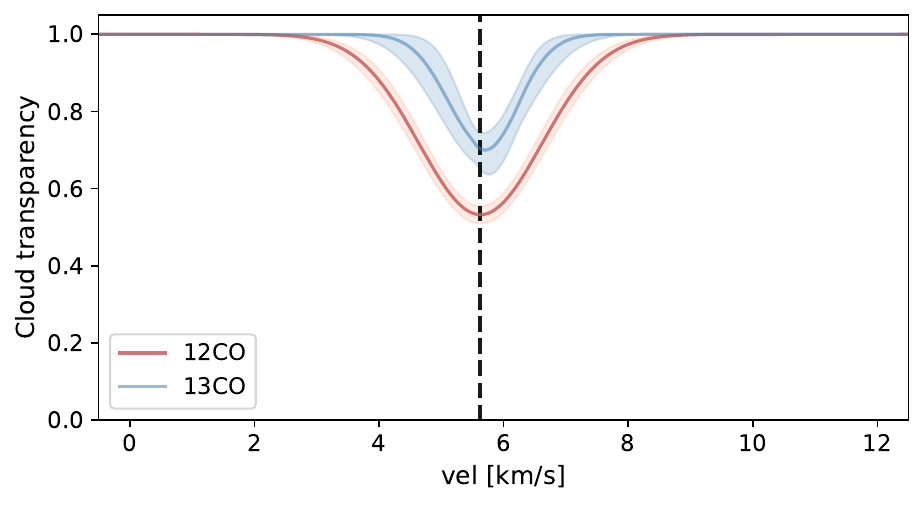}\\ \vspace{-0.2cm}
   \caption{Cloud transparency for $^{12}$CO and $^{13}$CO emission. The shaded regions show the $1\sigma$ confidence levels. The systemic velocity from Table~\ref{tab:mcmc_results} is shown with a dashed line. }
   \label{fig:app:cloud}
\end{figure}

\begin{figure*}[t]
\centering
    \includegraphics[width=0.9\textwidth]{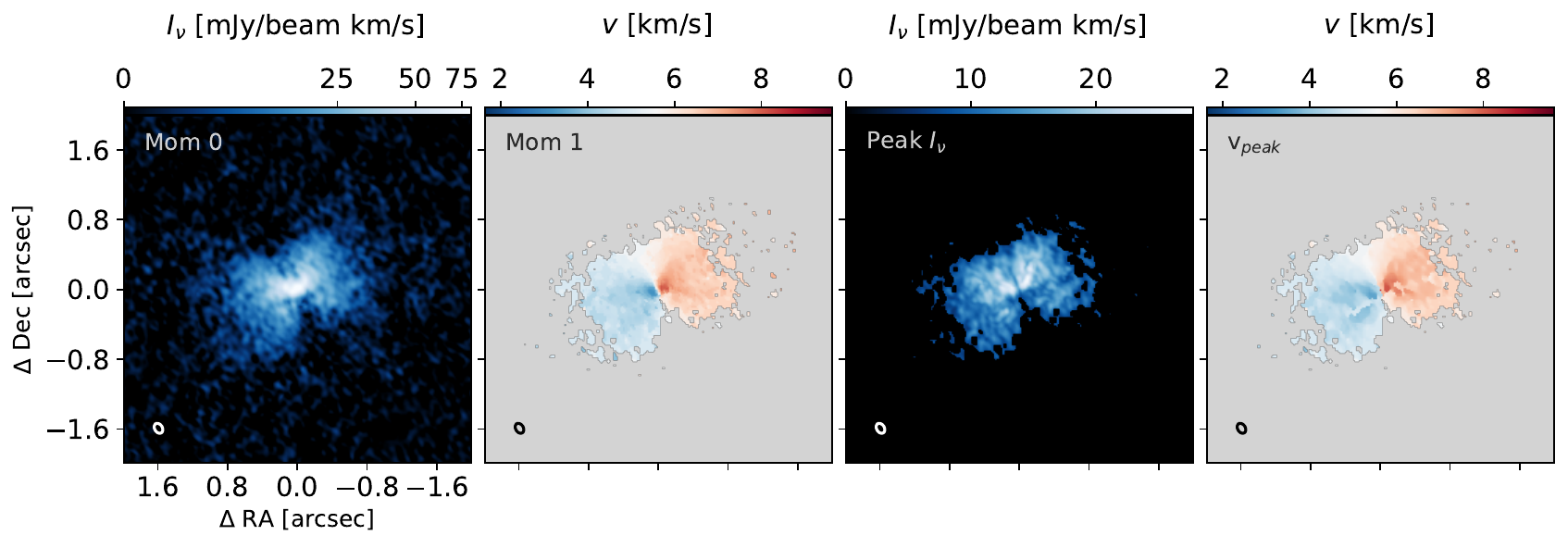}
    \includegraphics[width=0.9\textwidth]{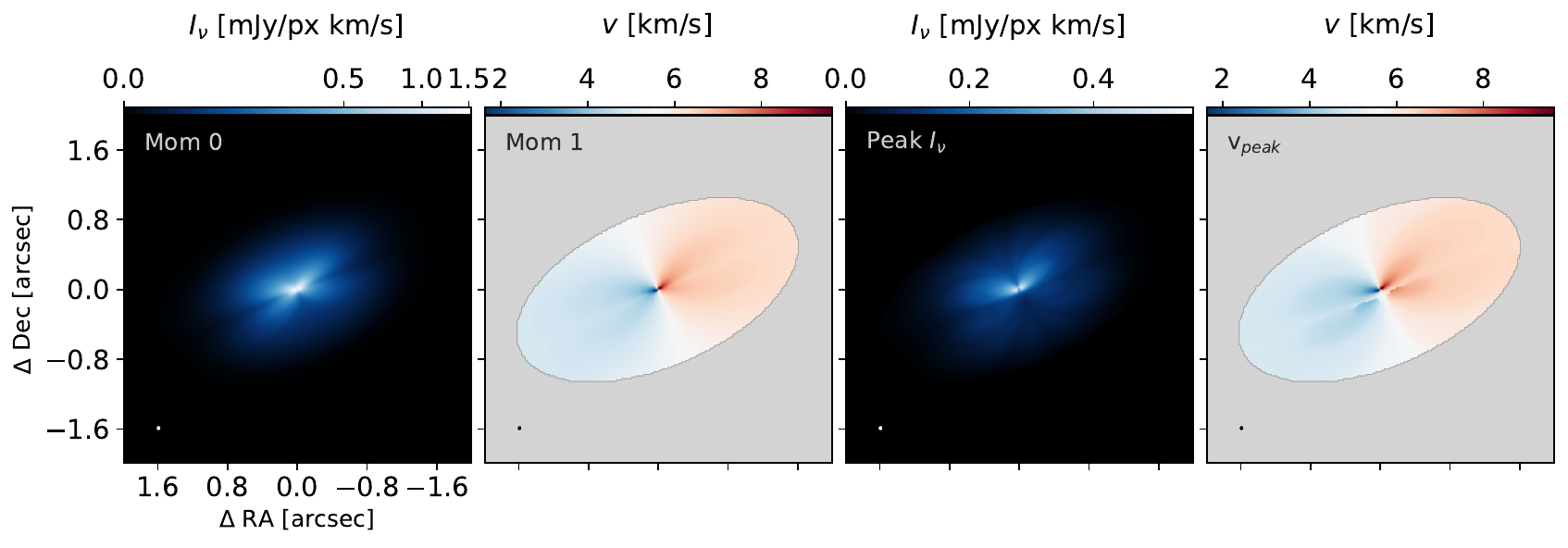}
    \includegraphics[width=0.9\textwidth]{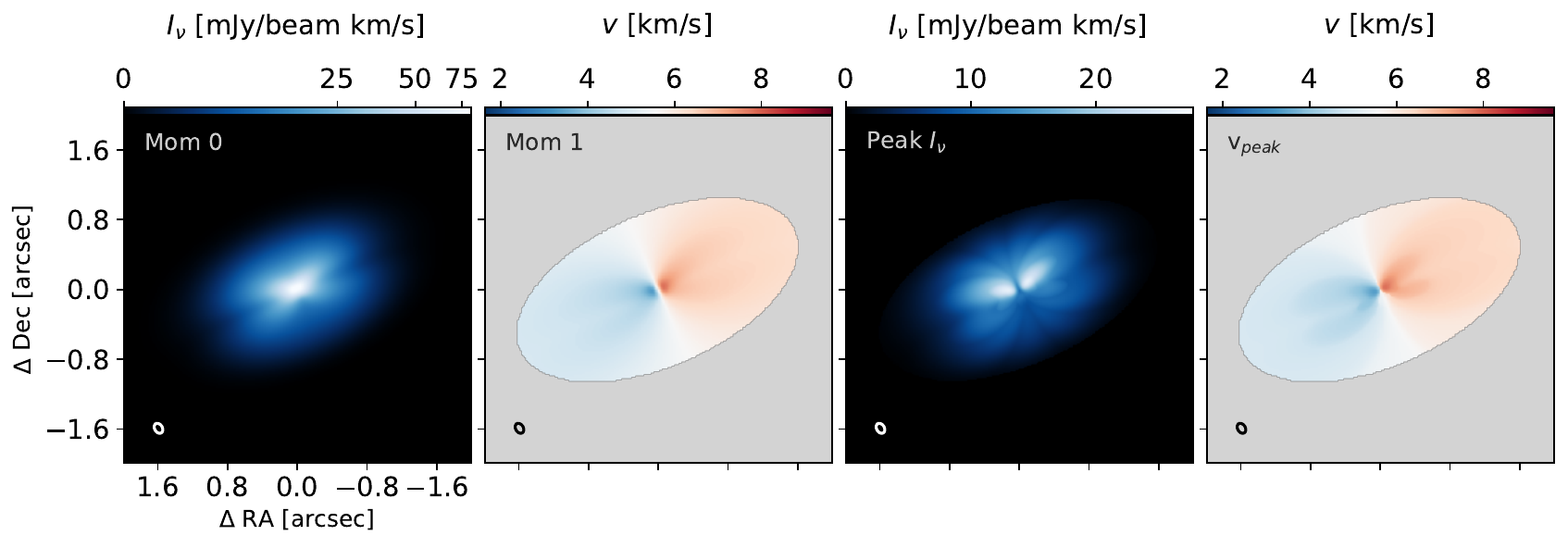}
    \includegraphics[width=0.9\textwidth]{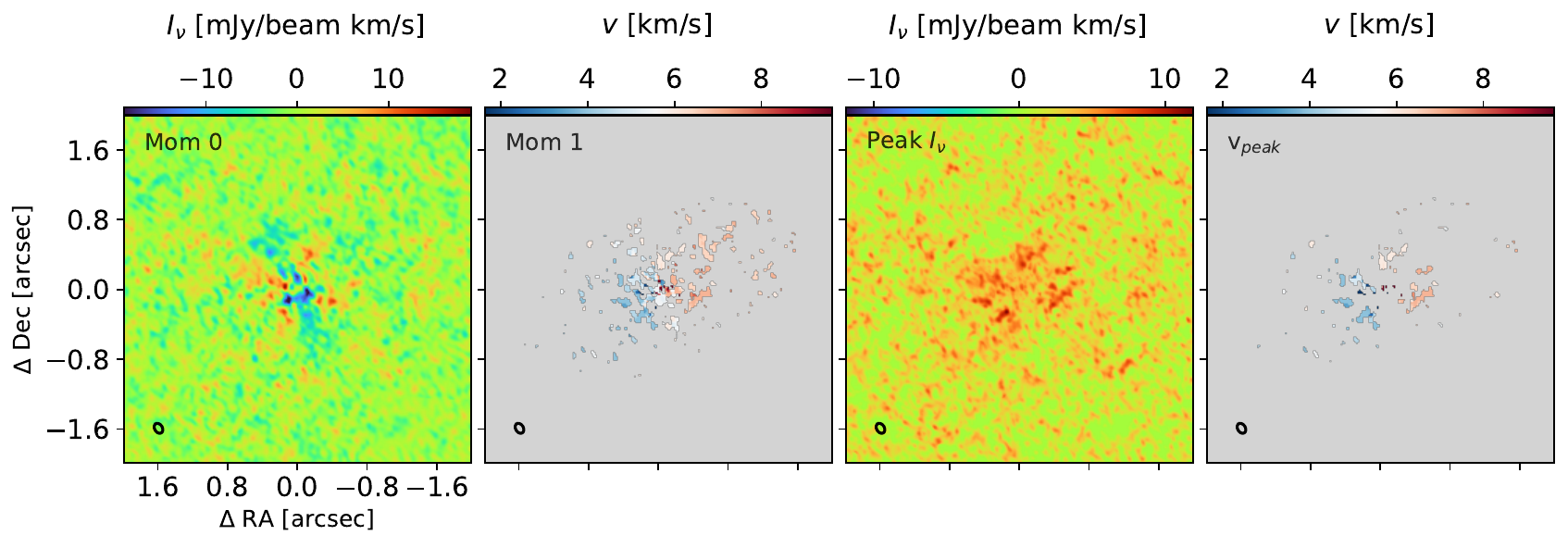}
    \caption{ Moment maps of the $^{12}$CO emission. From top to bottom row: Moment maps of the ALMA observation, the best model with a nominal pixel resolution, the best model after beam convolution, and the residual images. All moment maps where generated using a Keplerian mask. }
   \label{fig:app:moment_maps}
\end{figure*}

The Figures showing the modeled cloud transparency and the moment maps for the $^{12}$CO observation and model are shown in Figs.~\ref{fig:app:cloud} and \ref{fig:app:moment_maps}, respectively. The best parameters of the $^{12}$CO and $^{13}$CO models are shown in Table~\ref{tab:mcmc_results}.

\begin{table*}[t]
\caption{Results of the MCMC fitting for the $^{12}$CO and $^{13}$CO emission of MHO\,6. The uncertainties are measured from the 16th and 84th percentiles of each distribution. The fluxes ($F_{sky}$, $F_{corr}$) and the radii ($R_{68\%}$, $R_{90\%}$, $R_{95\%}$) are calculated from the models, and they are not free parameters.}
\centering
\begin{tabular}{ c|c|c|c||c|c|c|c } 
  \hline
  \hline
\noalign{\smallskip}
Parameter            & $^{12}$CO            & $^{13}$CO              & units & Parameter            & $^{12}$CO            & $^{13}$CO              & units \\
\noalign{\smallskip}
  \hline
\noalign{\smallskip}
 $\delta_{\rm{RA}}$  & $6.3_{-1.5}^{+1.0}$   & $7.8_{-3.7}^{+3.3}$   & mas  &  $I_b$       & $0.36_{-0.03}^{+0.01}$    & $0.17_{-0.04}^{+0.02}$    & mJy/px \\
 $\delta_{\rm{Dec}}$ & $5.2_{-1.8}^{+1.3}$   & $1.6_{-4.8}^{+5.1}$   & mas  &  $\alpha_b$  & $0.01_{-0.01}^{+0.01}$    & $0.04_{-0.4}^{+0.23}$     & - \\
 inc                 & $59.1_{-0.5}^{+2.8}$  & $59.6_{-1.4}^{+1.7}$  & deg  &  $\beta_b$   & $2.95_{-0.06}^{+0.38}$    & $1.92_{-0.06}^{+0.69}$    & - \\
 PA                  & $293.5_{-0.3}^{+0.2}$ & $293.8_{-0.9}^{+0.5}$ & deg  &              &                           &                           &   \\
\noalign{\smallskip}
  \hline
\noalign{\smallskip}
 $M_\star$        & $0.164_{-0.003}^{+0.003}$ & $0.171_{-0.010}^{+0.009}$  & $M_\odot$   & $\delta v_0$     & $457.9_{-7.4}^{+6.6}$     & $780.9_{-42.4}^{+48.3}$    & m\,s$^{-1}$ \\
 VLSR             & $5627.0_{-3.6}^{+3.8}$    & $5690.9_{-13.4}^{+13.4}$   & m\,s$^{-1}$ & $q$     & $0.40_{-0.02}^{+0.02}$    & $0.15_{-0.05}^{+0.05}$     & - \\
                  &                           &                            &             & $\delta v_{max}$ & $<1371.4_{-44.5}^{+430.8}$ & $<1466.7_{-309.5}^{+359.8}$ & m\,s$^{-1}$ \\
\noalign{\smallskip}
  \hline
\noalign{\smallskip}
 $r_z$            & $49.5_{-6.9}^{+46.8}$     & $105.1_{-4.6}^{+138.3}$    & au     & $cl_{v0}$        & $2.6_{-14.2}^{+14.4}$     & $-6.1_{-72.6}^{+81.8}$    & m\,s$^{-1}$ \\
 $r_I$            & $152.1_{-0.1}^{+3.5}$     & $98.1_{-2.05}^{+10.0}$     & au     & $cl_\sigma$      & $986.2_{-70.9}^{+77.1}$   & $563.3_{-151.4}^{+180.9}$ & m\,s$^{-1}$ \\
 $r_{max}$        & $<839.5_{-450.1}^{+49.5}$ & $<578.3_{-287.0}^{+283.8}$ & au     & $cl_a$           & $0.47_{-0.02}^{+0.02}$    & $0.30_{-0.04}^{+0.07}$    & - \\
 $r_{cav}$        & $1.6_{-0.9}^{+1.0}$       & $4.2_{-1.2}^{+0.5}$        & au     &                  &                           &                           & - \\
\noalign{\smallskip}
  \hline
\noalign{\smallskip}
 $z_0$            & $205.2_{-60.9}^{+59.3}$   & $24.7_{-3.7}^{+23.9}$      & au     & $F_{sky}$  & $6.56_{-0.05}^{+0.05}$  & $1.42_{-0.03}^{+0.04}$ & mJy km\,s$^{-1}$ \\
 $\psi$         & $1.56_{-0.13}^{+0.10}$    & $0.60_{-0.09}^{+0.15}$     & -      & $F_{corr}$ & $9.68_{-0.38}^{+0.46}$  & $1.59_{-0.07}^{+0.10}$ & mJy km\,s$^{-1}$\\
 $\phi$         & $0.87_{-0.10}^{+0.15}$    & $19.15_{-18.08}^{+27.57}$  & -      &            &                         &                        &     \\
\noalign{\smallskip}
  \hline
\noalign{\smallskip}
 $I_f$            & $0.42_{-0.03}^{+0.02}$    & $0.09_{-0.01}^{+0.02}$     & mJy/px & $R_{68\%}$ & $131.2_{-1.8}^{+1.9}$ & $ 72.4_{-4.3}^{+3.8}$   & au \\
 $\alpha_f$       & $0.32_{-0.01}^{+0.03}$    & $0.80_{-0.11}^{+0.05}$     & -      & $R_{90\%}$ & $182.3_{-3.0}^{+3.0}$ & $ 94.6_{-8.2}^{+8.7}$   & au \\
 $\beta_f$        & $2.58_{-0.11}^{+0.13}$    & $15.74_{-11.26}^{+3.76}$   & -      & $R_{95\%}$ & $203.6_{-3.3}^{+3.2}$ & $102.2_{-10.7}^{+11.9}$ & au \\
\noalign{\smallskip}
  \hline
  \hline
\end{tabular}
\label{tab:mcmc_results}
\end{table*}

\section{Alternative combination of emitting layers}\label{app:sec:appendix_coord}

\begin{figure*}
 \centering
        \includegraphics[width=0.9\textwidth]{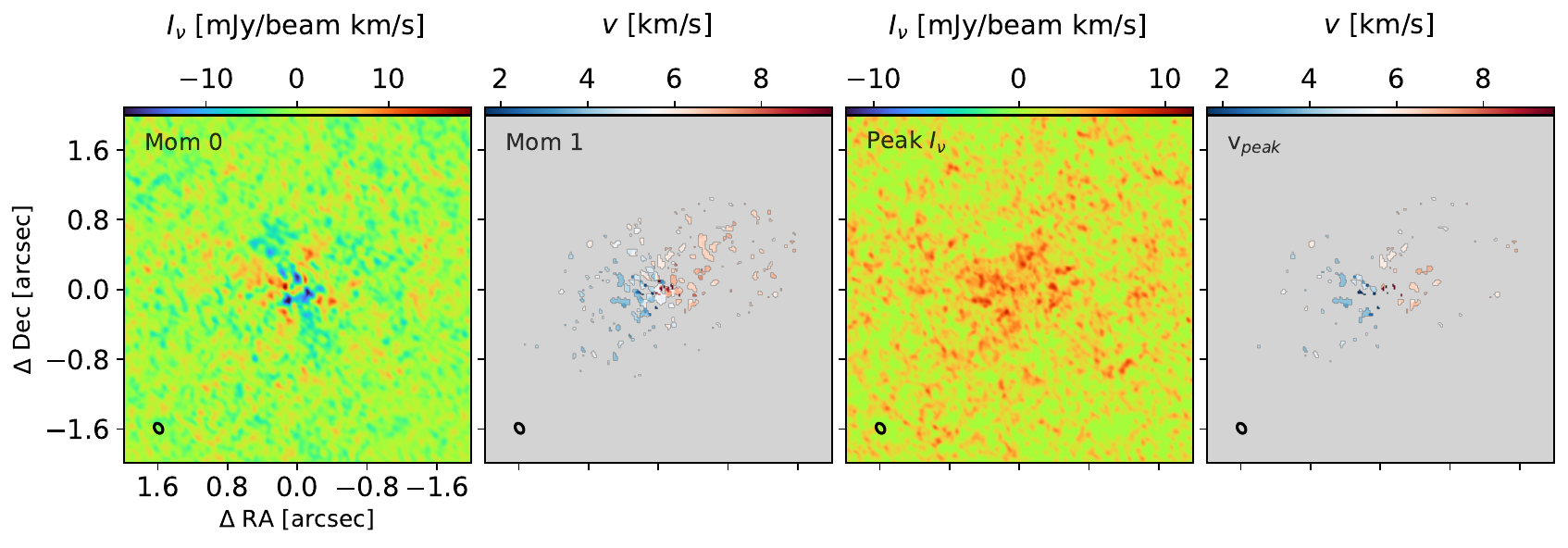}\\ \vspace{-0.2cm}
   \caption{Moment maps of the residuals from the best model where the front and back layers where combined with Eq.~\ref{app:eq:I_k}. All moments where generated using a Keplerian mask.}
   \label{fig:app:momres}
\end{figure*}

Alternatively to Eq.~\ref{eq:I_k}, we experimented using a smooth transition from $I_{f,k}$ to $I_{b,k}$ with a power law weighting:
\begin{equation}
\centering
    I_k = 
    \left\{
    \begin{aligned}
        I_{f,k}\,\text{,} & \,\,\,\,\,\text{for } I_{f,k}\geq \mu \\
        I_{f,k} + \epsilon I_{b,k}\,\text{,} & \,\,\,\,\,\text{for } I_{f,k}< \mu\\
    \end{aligned}
    \right.\text{,}
    \label{app:eq:I_k}
\end{equation}
\noindent where $\mu$ is the threshold at which the emission of the front layer stops being completely optically thick, and $\beta$ is a function of range $[0,1]$ calculated in the region of interest as:
\begin{equation}
\centering
    \epsilon = 1 - \left( \frac{I_{f,k}}{\mu} \right)^\gamma\text{,}
\end{equation}
\noindent thus, the combination is described with two free parameters, $\mu$ and $\gamma$. A fit to the $^{12}$CO data was done using this $(\mu,\gamma)$ approximation to combine the layers, with consistent results for $z_f$, $I_f$ and $I_b$ compared to Eq.~\ref{eq:I_k}. The best results were $\mu=0.355\pm0.04$\,mJy/px and $\gamma=0.3\pm0.05$, but no significant change is seen in the residuals map, as shown in Fig.~\ref{fig:app:momres}. 

\section{Nesting images}\label{app:sec:nesting_images}

\begin{figure*}
 \centering
        \includegraphics[width=17.6cm]{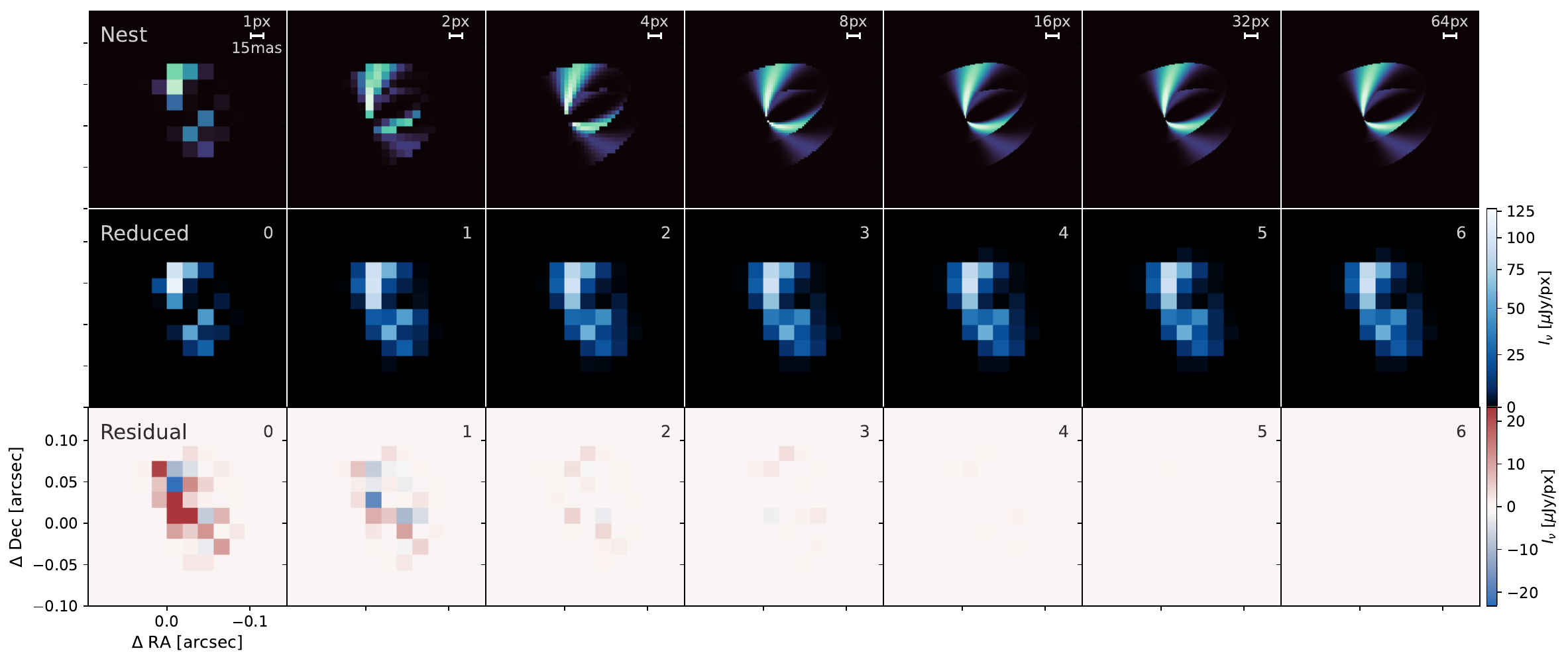}\\ \vspace{-0.2cm}
   \caption{For a given velocity channel, the same model is shown with different nesting levels. All the nested images are then integrated into a reduced image, with the same pixel size for all of them. The lower row shows the residual between the reduced image generated with $n=6$ and models with lower $n$.}
   \label{fig:app:nested_examples}
\end{figure*}

\begin{figure}
 \centering
        \includegraphics[width=7.5cm]{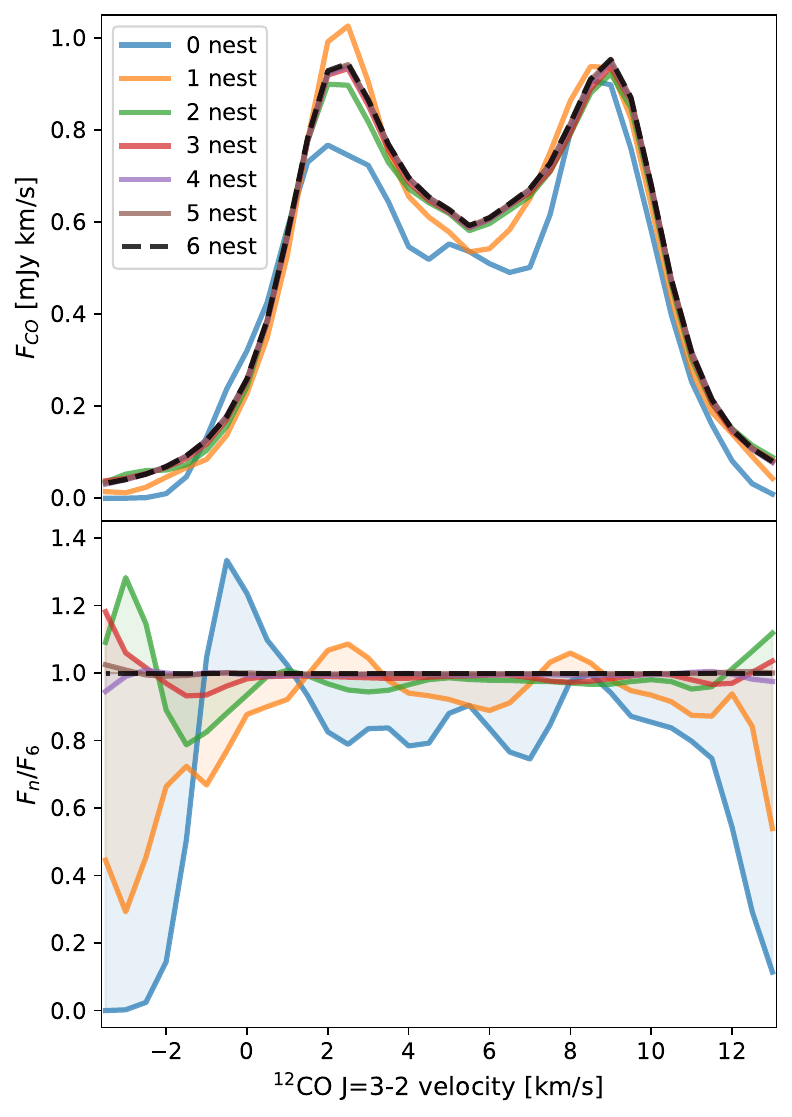}\\ \vspace{-0.2cm}
   \caption{Flux variations as a function of velocity, for a disk of 10\,au in size with the same model parameters as those from the $^{12}$CO of MHO\,6. \textbf{Upper panel}: Total flux as a function of velocity for different nesting levels. \textbf{Lower panel}: Flux ratio between a model with $n=6$ and models with lower nesting level. }
   \label{fig:app:nest_diff}
\end{figure}

\begin{figure}
 \centering
        \includegraphics[width=7.5cm]{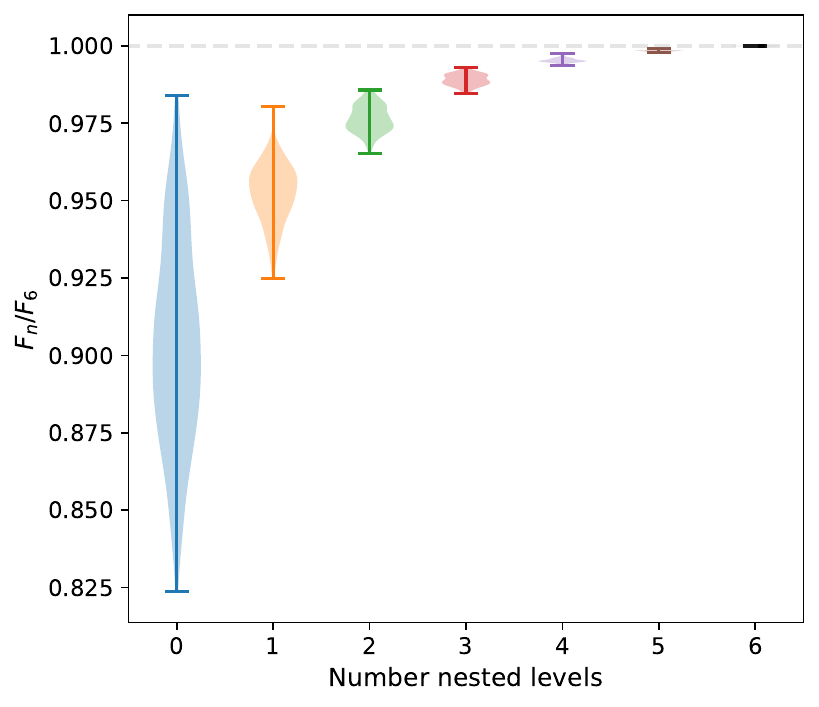}\\ \vspace{-0.2cm}
   \caption{Distribution of flux ratio between models with different levels of nesting and a model with a nesting of $n=6$ as a reference. For each nested level, 256 models where generated by shifting the center of the disk by 1\,mas in the RA and Dec grid, covering the full extent of one nominal pixel. For each nested level the width of the shape is representative of the density distribution of the sample. }
   \label{fig:app:violin_diff}
\end{figure}

The nesting image process is used to create model images with higher flux fidelity for the inner disk region, oversampling the nominal images by $4^n$, where $n$ is the level of nesting. In order to compare the flux difference between models with the same parameters but different nesting levels, we generate model images for a disk of 10\,au in radius with similar model parameters to those of MHO\,6 $^{12}$CO emission. The nominal pixel size is set to $15\,$mas, as in the $^{12}$CO fit. 

We generated models from $n=0$ (no nesting) to $n=6$ (oversample each nominal pixel with $4^6$ pixels), and took the model images with $n=6$ as the reference for true flux. In Fig.~\ref{fig:app:nested_examples}, we show the residuals between each reduced image relative to the $n=6$ reduced image for a single channel. In Fig.~\ref{fig:app:nest_diff}, we show the total flux difference between $n=6$ and lower levels of nesting as a function of velocity. The models with $n=0$ and $n=1$ show the largest deviations from the true flux. In the lower panel of Fig.~\ref{fig:app:nest_diff}, the proportion between the flux of each model is compared to that of $n=6$, showing that even with $n=2$ or $n=3$, the channels with very faint emission could have discrepancies as large as 30\% at high velocities. Thus, the nesting level has to be chosen so that those variations do not dominate the noise during the fitting process. 

In order to investigate the impact of small spatial variations of the disk center relative to the pixel grid, we ran the same model with levels of nesting from $n=0$ to $n=6$, changing the disk center by 1\,mas in RA and Dec until a full 15\,mas offset was completed (the pixel size in the nominal image). This gave us a grid of 256 models per level of nesting. The flux variation is shown in Fig.~\ref{fig:app:violin_diff}, where it can be seen that spatial offsets of the disk center can change the total flux of the inner 10\,au by a factor of $\sim15\%$ when no nesting is being applied. Increasing $n$ has an asymptotic improvement over the flux conservation.

\end{appendix}

\end{document}